\documentclass[english,twocolumn]{article}
\usepackage[T1]{fontenc}
\usepackage[latin9]{inputenc}
\usepackage{geometry}
\usepackage{graphicx}
\usepackage{esint}

\makeatletter

\providecommand{\tabularnewline}{\\}

\newcommand{\lyxaddress}[1]{
\par {\raggedright #1
\vspace{1.4em}
\noindent\par}
}

\makeatother

\usepackage{babel}
\begin{document}

\title{A Graphical Approach to Radio Frequency Quadrupole Design }

\author{G. Turemen$^{a}$, G. Unel$^{b}$, B. Yasatekin$^{a}$\thanks{byasatekin@ankara.edu.tr}}

\maketitle

\lyxaddress{$^{a}$Ankara University, Department of Physics, Graduate School
of Natural And Applied Sciences, Ankara, TURKEY.\\
$^{b}$University of California at Irvine, Department of Physics and
Astronomy, Irvine, USA}
\begin{abstract}
\indent 

The design of a radio frequency quadrupole, an important section of all ion accelerators, and the calculation of its beam dynamics properties can be achieved using the existing computational tools. These programs, originally designed in 1980s, show effects of aging in their user interfaces and in their output. The authors believe there is room for improvement in both design techniques using a graphical approach and in the amount of analytical calculations before going into CPU burning finite element analysis techniques. Additionally an emphasis on the graphical method of controlling the evolution of the relevant parameters using the drag-to-change paradigm is bound to be beneficial to the designer. A computer code, named DEMIRCI, has been written
in C$^{++}$ to demonstrate these ideas. This tool has been used in the design of Turkish Atomic Energy Authority (TAEK)'s 1.5~MeV proton beamline at Saraykoy Nuclear Research and Training Center (SANAEM). DEMIRCI starts with a simple analytical model, calculates the RFQ behaviour and produces 3D design files that can be fed to a milling machine. The paper discusses the experience gained during design process of SANAEM Project Prometheus (SPP) RFQ and underlines some of DEMIRCI's capabilities.
\end{abstract}

\section{Introduction}

\indent

The Radio Frequency Quadrupoles (RFQ) are in use at the low beta sections of all modern ion accelerators since their invention by Kapchinsky and Teplyakov in late 1970s~\cite{TepilKap}. For light ions such as $H^{+}$ and $H^{-}$ operating in 300-400~MHz range, the RFQ type of choice is the, so called, \textquotedbl{}4-vane\textquotedbl{} RFQ \cite{wangler}. The design of a 4-vane RFQ, which is the focus of this paper, and its manufacture require precise calculation of the relevant parameters, a good understanding of the materials and high precision machining~\cite{staples}. In fact, this accelerating structure is nothing but a body (cylindrical, square or octagonal vessel) containing four carefully crafted modulated vanes, two vertical and two horizontal symmetrically distributed along the beam axis. The high precision modulation requirement on the vanes can be met by the computerized milling tools, i.e. CNC machines. However, the art of designing an efficient RFQ and the study of its beam dynamics properties necessitate repetitive lengthy calculations:~An ideal task for computers. Although the two of commercially available programs~\cite{Lidos}\cite{Parmteq}, profit from years of experience in accelerator building, the main design ideas and especially the user interaction components can benefit from modern tools and concepts. Additionally, the commonly used Unix-like environment provided by Linux and OSX workstations does not have access directly to these two Microsoft Windows specific software packages.

A new project in the form of a computer code, written in C$^{++}$, called DEMIRCI\footnote{meaning ``blacksmith'' in Turkish}~\cite{demirci} is started to explore the potential of the modern concepts such as object oriented programming and ROOT environment~\cite{root}. This tool helps the designer to create an RFQ model which would achieve certain goals such as a final target energy or a fixed total accelerator length in a fully graphical environment. It calculates a large number of design and beam dynamics parameters such as energy at the end of the cavity, power dissipation and cavity quality factor for each cell. It also allows the designer to visualize a large set of parameters change along the RFQ. Another property of this tool is the interoperability with similar software in the field, either directly using the user interface or by simple exchange of
plain text files. 

This paper focuses on the algorithmic side of the project omitting all the installation details and instructions on how to use the actual code. These details are described elsewhere~\cite{demirci}. However it might be interesting for the reader to know that the project has been used successfully in multiple Unix-like environments such as OSX, Scientific Linux and Ubuntu Linux. At the writing of this note the project source code is under further development and is not open to general public, however a binary at the current state for most common platforms can be obtained by contacting the authors~\cite{get_code}.

\section{Design Procedures }

\indent 

The classical procedure used in designing 4-vane RFQs has been around since LANL designed the first proof of principle (PoP) device. This procedure is known as the ``LANL Four Section Procedure (FSP)'' method~\cite{fsp}. According to this method, the RFQ is divided into 4 sections named as radial matching section (RMS), shaper section, gentle buncher section and acceleration section. After the steady state beam at the entrance of the RFQ is matched to time dependent electric field in the RMS, an RF bucket is formed in the shaper to prepare the beam for the gentle buncher. While the beam is being bunched, the space charge effect is also attempted to be reduced concurrently. After the beam becomes bunched, it is accelerated to the final energy
at the accelerator section. 

\textit{\emph{The potential between the electrodes of a single RFQ cell is given}}\cite{Biscari}\textit{\emph{ by:}}
\begin{eqnarray}
U(r,\theta,z) & = & \frac{V}{2}\Bigg[\sum_{m=1}^{\infty}A_{0m}(\frac{r}{r_{0}})^{2m}\cos(2m\theta)\label{eq:generic-potential}\\
 & + & \sum_{m=0}^{\infty}\sum_{n=1}^{\infty}A_{nm}I_{2m}(nkr)\cos(2m\theta)\cos(nkz)\Bigg]\nonumber 
\end{eqnarray}
where $r$ and $\theta$ are cylindrical coordinates for which $z$ represents the beam direction, $V$ is the
amplitude of the inter-vane voltage, $k$
is the wave parameter given by $k\equiv2\pi/\lambda\beta$, with $\lambda$ being the RF wavelength and $\beta$ being the speed of the ion relative to the speed of light. Also, $r_{0}$ is mean aperture of the vanes, $I_{2m}$ is the modified Bessel function of order 2m and the $A_{nm}$ are the multipole coefficients whose values depend on the vane geometry. 

Kapchinsky and Teplyakov argued that for practical purposes the above potential can be approximated only by the lowest order terms in the sums (hence the name ``2-term potential'') to calculate the EM fields around the tips of the electrodes i.e. near the beam axis. More recently, modern tools have added few more terms to this initial approximation, in fact LANL design software uses the eight lowest order terms of Equation~(\ref{eq:generic-potential}) to characterize the EM fields in the presence of the modulated vanes. \textit{\emph{The other commercially available software, LIDOS, gives users the possibility to design RFQs
in three steps; first the main parameters are defined and design optimizations are made, then accurate RF fields calculations are made with a multipole expansion of the equation~\ref{eq:generic-potential} and finally beam simulations are performed to understand the beam dynamics effects.}}

\subsection{New design procedure}

\indent

\begin{figure}[!htb]
\begin{centering}
\includegraphics[width=1\columnwidth]{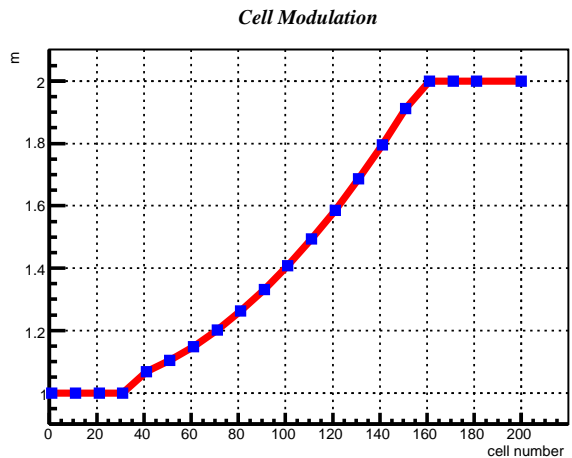}
\par\end{centering}
\caption{Cell modulation ($m$) versus cell number. $m$ is a typical RFQ parameter
which changes with cell number. The meaning of blue squares and red
line is explained in the text. \label{fig:Cell-modulation}}
\end{figure}

The parameters needed to define an RFQ can be divided into two categories:~the ones which can be a function of RFQ length and the ones which are constant for a given RFQ. The resonant frequency ($f$), the initial ion energy ($E_{in}$), the input beam current ($I$) and the braveness factor (in terms of the Kilpatrick value) can be cited as examples to the latter. The four parameter vectors falling into the first category are:~the synchronous phase ($\phi$), the cell modulation ($m$), the minimum bore radius ($a$) and the inter-vane voltage ($V$). 
This last one, together with $r_{0}/\rho$ where $r_{0}$ is mean aperture of the vanes and $\rho$ is the curvature (tip radius) of the electrodes, 
could be kept constant along the RFQ length to simplify the design and manufacture. 
In case of DEMIRCI, a typical parameter's variation along the RFQ can be seen in Fig.~\ref{fig:Cell-modulation}. The points represented by the blue squares in Fig.~\ref{fig:Cell-modulation} are the so called
\textquotedbl{}reference cells\textquotedbl{} for which the values of the four key parameters are defined by the designer. In this particular example, Fig.~\ref{fig:Cell-modulation} shows 20 reference cells for an RFQ of 200 cells in total. The values of the parameters at the cells in between the reference ones are obtained by interpolation assuming a simple linear function. 

The number of reference cells and the total number of RFQ cells are all user defined variables. The designer might choose to define the values of the parameters for each cell, or to simply define boundary conditions for different regions of the RFQ and let the interpolation function do the rest. As a safety check, the software library ensures the monotonic increase of the
reference cell numbers. Therefore a new design can be made in a pure graphical way, by simply relocating individual reference cells by using the mouse pointer e. g. to change the shape of the synchronous phase curve or by moving a complete curve, e.g. to increase the inter-vane voltage which is a constant in simple designs. This new paradigm allows quick testing of various design ideas concerning the four critical parameter vectors:~$\phi$, $m$, $a$ and $V$. Although the non-graphical user interface, i.e. the command line, also exists and it could be more adequate for parameter scan studies, the graphical method has proven itself to be both more intuitive and more pedagogical for the new designers. 

Once the reference curves and the other parameters are selected, the designer can simply do the interpolation and calculate all the relevant functions for each cell. The evolution of the ion beam starts with cell 0 and progresses through all cells by calculating all the variables of interest. 
A few simple formulas leading to the calculation of the length of the $n^{th}$ cell $L_{n}$, hold at voltage $V_n$ and synchronous phase $\phi_{n}$, the acceleration efficiency at $n^{th}$ cell considering 2-term potential, $A_{n}$, and the total energy ($E_{n}$) of the ions of mass $m_{p}$, charge $q$ and speed of light $c$, after the $n^{th}$ cell, 
are reproduced below to give the reader an overview of the type of repetitive calculations needed to design an RFQ: 
\begin{eqnarray}
\beta_{n} & = & \sqrt{1-(m_{p}c^{2}/E_{n-1})^{2}}\label{eq:simple_example}\\
L_{n} & = & \beta_{n}\lambda/2\nonumber \\
k_{n} & = & 2\pi/(\lambda\beta_{n})\nonumber \\
A_{n} & = & (m_{n}^{2}-1)/(m_{n}^{2}I_{0}[k_{n}a_{n}]+I_{0}[k_{n}a_{n}m_{n}])\nonumber \\
E_{n} & = & E_{n-1}+\frac{1}{4}q\pi A_{n}V_{n}\cos(\phi_{n})\quad.\nonumber 
\end{eqnarray}
 
\begin{figure}[!htb]
\includegraphics[width=9cm,height=5cm]{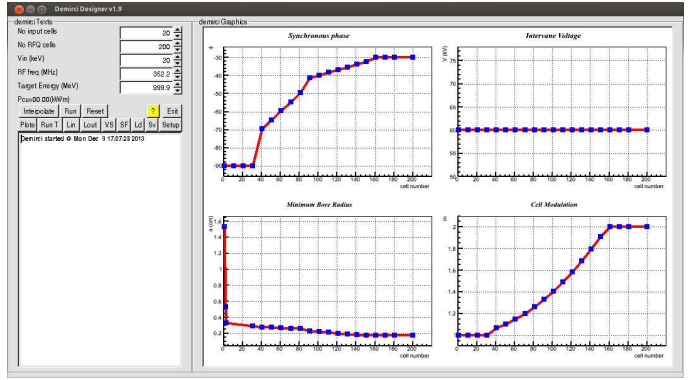}\caption{The graphical interface of DEMIRCI. \label{fig:demirwindow}}
\end{figure}

At the end of the calculations, few important parameters are presented to the designer:~These are the final ion beam energy, the RFQ total length which is the sum of all cell lengths, the time needed for the ions to travel the RFQ cavity, the maximum EM power needed by the RFQ and its quality factor, $Q$. The details of the RF power and quality factor calculations will be discussed in the next section. Remaining parameters can be optimized by specifying a goal such as a desired output beam energy. The number of RFQ cells can be changed from the default value of 200 to allow the design of longer RFQ cavities. A shorter design is simpler to obtain by setting a target value for the exit ion energy. All the scalar RFQ parameters can be tuned using the number entry boxes at the upper right side of the designer window which can be seen in Fig.~\ref{fig:demirwindow}. 

\begin{figure}[!htb]
\begin{centering}
\includegraphics[width=0.9\columnwidth]{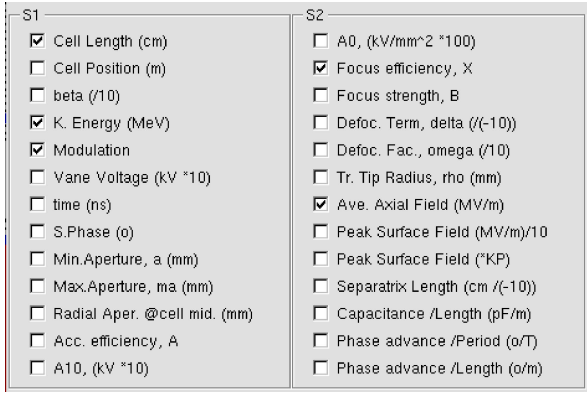}
\par\end{centering}
\caption{A screenshot from DEMIRCI:~Drawable parameter selection.\label{fig:Plottables}}
\end{figure}

Additionally, there are checks along the calculations which inform the user that a critical parameter is overrun. For example, the designer is warned if the inter-vane voltage is too 
high and induces an electric field above the preselected Kilpatrick limit. Since DEMIRCI receives from the designer the braveness factor (between 1 and 2) 
and the operation frequency ($f$ in MHz), it calculates internally the maximum allowed electric field ($E_{max}$) by inverting Kilpatrick's empirical limit 
formula: $f=1.643*E^2*e^{\frac{-8.5}{E}}$ and multiplying the resulting field $E$ by the braveness factor~\cite{staples}. 
Then, the peak electric field at the electrode surface ($E^s_n$) at $n^{th}$ cell is calculated and checked against $E_{max}$, using inter-vane voltage and bore radius 
values in that cell: $ E^s _n=\kappa  V_n/ r_n$, where $\kappa$ is selected as 1.28 to ensure compatibility with CST and Superfish results ~\cite{cst, superfish}.
If such a warning is given, it is up to the designer to check the details of the problem by plotting the parameter in question and to solve it by re-optimizing the RFQ. Building the RFQ designer software on top of the pre-existing ROOT libraries provides all the non-essential but necessary functionality, such as defining the parameters of interest, loading and saving of
the configuration and output files. Additionally, all the graphics routines in DEMIRCI are based on the ROOT libraries. This design decision allows a robust, mature and multi-platform GUI experience for the designer. A section of the user interface dealing with parameter selection is shown in Fig.~\ref{fig:Plottables}. DEMIRCI provides easy plotting of the evolution of the relevant parameters along the RFQ. The graphical results can also be easily customized (such as the formatting of the curves or the addition of a gridline) according to the taste of the user. 

\subsection{Power and quality factor calculations}
\indent
RF power requirements and the quality of the final accelerating cavity are essential quantities to measure the realizability of the designed device. In DEMIRCI, the RF power dissipation on the cavity walls is calculated using the lumped circuit model assuming a Cu structure~\cite{LumpedCircuit}. This assumption is expected to be valid for both Oxygen Free Electric Copper~(OFE Cu) cavities and cavities made from other materials and electro-coated with Cu:~The skin depth at radio frequencies is less than 10~$\mu$m, which is much less than typical coating thickness validating the approach. The relevant quantity to consider is the power loss per unit length on the RFQ wall surfaces. It is denoted as $P_{\ell}^{s}$ and defined as~\cite{cst}:
\begin{equation}
P_{\ell}^{s} \equiv \frac{1}{2}\sqrt{\frac{f\pi\mu_0 }{\sigma}} \frac{1}{\mu_0 ^2} \int \left| {\bf B}\right|^{2}dp  \quad,
\label{Ps_defr}
\end{equation}
\noindent where $\mu_{0}$ is the magnetic constant, ${\bf B} $ is the magnetic field vector and $\sigma$ is the conductivity of the wall material (usually Cu).
The operating frequency ($f$) is defined using $w=2\pi f$ where  $\omega$ is the resonant angular frequency. 
The function is to be integrated over the perimeter ($dp$) of the surface at a given RFQ slice along its length. The perimeter and the surface of the RFQ slice
can be calculated assuming that each quadrant can be approximated by a square of length $r$ and three quarters of a circle of radius $r$. This simplified geometry is also known as "clover leave" 4-vane resonator. 
Therefore the surface area (S) of a section with four quadrants and its perimeter length (P) become :
\begin{eqnarray}
 S &=& r^2(4+3 \pi) \quad,  \label{SandP} \\
 P &=& 2r(4+3\pi) \quad. \nonumber
\end{eqnarray}
The resonance condition of this capacitor-inductor cavity with the above geometry can be used to calculate the relation between $r$ and various other quantities~\cite{LumpedCircuit}:
\begin{equation}
r^2=\frac{16}{4+3 \pi}\frac{1}{\mu_0 C_{\ell} w^2} \quad, \label{rdef} 
\end{equation}
\noindent where $C_{\ell}$ is the capacitance per unit length.
If the specific cavity design is known, an adapted version of the resonance equation can be used to calculate the  $C_{\ell}$  value.
However DEMIRCI uses an estimation for this last quantity, a constant around 120~pF/m, to determine $r$.
In fact, a better approximation can be found in the literature:~$C_{\ell}=48\times10^{-12}(r_{0}/\lambda)^{-1/6}$ for $0.002\leq r_{0}/\lambda<0.008$~\cite{LumpedCircuit} where $\lambda$ is the RF wavelength and $r_{0}$ is radial aperture. 
Therefore the integral in eq. (\ref{Ps_defr}) simplifies to:
\begin{equation}
P_{\ell}^{s}=\sqrt{\frac{4+3\pi}{32\sigma}}(\omega C_{\ell})^{3/2}V^{2}\quad,
\end{equation}
\noindent where V is the inter-vane voltage as before. 
The actual consumed power by a specific machine is the sum of surface losses and power transferred to the beam ($P^{beam}$) which
depends on the beam current ($I^{beam}$) and its final energy ($E^{beam}$) :~$P^{tot}=P^{s}+P^{beam}$ where
\begin{equation}
P^{beam}=I^{beam} E^{beam}\quad.
\end{equation}
\noindent In order to calculate another important RFQ property, namely the quality factor of the cavity denoted as $Q$, 
the  magnetic  component of the stored energy per unit length ($U_{\ell}^{B}$) is to be calculated \cite{LumpedCircuit}:
\begin{equation}
U_{\ell}^{B} \equiv \frac{1}{2\mu_{0}}\int\left|\bf{B}\right|^{2}dS = \frac{V^2 C_\ell}{2} \quad.
\end{equation}
Therefore, without considering the effect of vane modulation, $Q$ is defined in literature as \cite{LumpedCircuit}:
\begin{equation}
Q=\omega\frac{U^{B}}{P^{s}}\quad, \label{oldQ}
\end{equation}
Here quantities without the subscript $\ell$ represent integration over the RFQ length. 
Recently, a simple consideration of the electric field in the $z$-direction has been investigated and it has been shown to modify the $Q$ value by about 2~\% as described in Appendix~A.
Finally, in all calculations the duty factor (\emph{d.f.}) is assumed to be 100\%, therefore the designer is expected to report 
the actual power values according to the \emph{d.f.} of the planned device.

\subsection{Machining the vanes}
\begin{figure}[!htb]
\includegraphics[width=1\columnwidth]{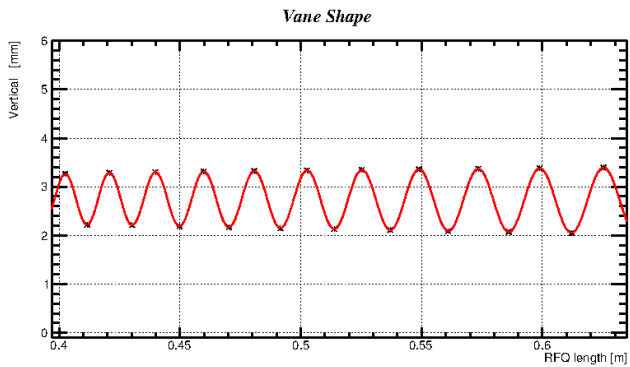}
\caption{Vane shape as calculated by DEMIRCI.\label{fig:Vane-Shape}}
\end{figure}

\indent
Vane shape calculation is the final output expected from such a software as it can be fed to an electromagnetic equation solver program for further analysis. The vane tip extrema position in radial direction is defined as part of the design process. The axial position (z) of each extrema is calculated by DEMIRCI library. An interpolation between consecutive extrema using generic sine function allows the definition of an equation of the curve that describes the distance from the beam axis to the electrode:
\begin{equation}
F(z)=c_0+c_1\sin(kz+\phi)\quad, \label{V_j}
\end{equation}
where the coefficients $c_i$, and $\phi$ can be solved for each cell. After the extraction of the extrema ($V_j$) at $j^{\mbox{th}}$ cell  from the DEMIRCI library, the vane shape related function firstly calculates the bore radius per cell, denoted as $c_0$ in equation \ref{V_j}. The other constant is defined as the difference of the extrema per cell and finally the phase can be extracted from the boundary condition that the sin function equals to unity at the beginning of each cell (k is the wave number given in equation \ref{V_j} ). Therefore one obtains:
\begin{eqnarray}
 c_0 &=& \frac{V_j + V_{j+1} }{2}  \quad , \endline
 c_1 &=& \frac{V_j - V_{j+1} }{2} \quad , \endline
\phi & = &  \pi/2 - kz             \quad .
\end{eqnarray}
Although the user can request the functions to be evaluated at any interval to produce a discreet version of the vane profile the typical precision requirement for mechanical productions can be met with a 10~$\mu m$ step size. In Fig.~\ref{fig:Vane-Shape} vane shape calculated by DEMIRCI is presented as a function of the RFQ length. Such an output file can be taken as input by a CAD program for 3D visualization and further processing.

\section{Prometheus RFQ as a Design Example}

\indent

The SPP RFQ, at TAEK's SANAEM, aims to gain the necessary knowledge and experience to construct a proton beamline needed for educational purposes. A Proof of Principle~(PoP) accelerator with modest requirements of achieving at least 1.5~MeV proton energy, with an average beam current of at least 1~$mA$ is under development. This PoP project has also the challenging goal of having the design and construction of the entire setup in Turkey within three years:~from its ion source up to the final diagnostic station, including its RF power supply. There are also two secondary goals of this project:~1) Training accelerator physicists and RF engineers on the job; 2) To encourage local industry in accelerator component construction. The design requirements of this machine can be found in Table~\ref{tab:SANAEM-Pop}. The input energy was selected to keep the RFQ short for a 1.5~MeV output energy and at the same time to satisfy the current requirements. The operating frequency was selected to be compatible with similar machines in Europe and therefore to benefit from the already available RF power supply market. Other parameters such as the inter-vane voltage, Kilpatrick value etc. were chosen to be adequate for a first time machine.

\begin{table}[!htb]
\caption{SPP RFQ Design Parameters\label{tab:SANAEM-Pop}}
 ~

\centering{}%
\begin{tabular}{c|c}
\textbf{Parameter} & \textbf{Value}\tabularnewline
\hline 
\hline 
$E_{in}$(keV) & 20\tabularnewline
\hline 
$E_{out}$ (MeV) & 1.5\tabularnewline
\hline 
$f$ (MHz) & 352.2\tabularnewline
\hline 
$V$ (kV) & 60\tabularnewline
\hline 
$I$ (mA) & >1\tabularnewline
\hline 
KP & 1.5\tabularnewline
\hline 
$r_{0}$ (mm) & 2.8\tabularnewline
\hline 
$\rho$ (mm) & 2.5\tabularnewline
\end{tabular}
\end{table}

\begin{figure}[!htb]
\begin{centering}
\includegraphics[width=1\columnwidth]{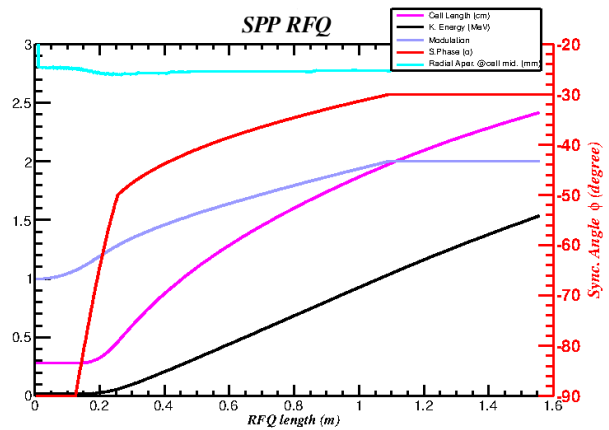}
\par\end{centering}
\caption{SPP RFQ beam dynamics parameters as calculated by DEMIRCI.\label{fig:SANAEM-POP-RFQ}}
\end{figure}

The calculation and visualization of the evolution of the most critical accelerator cavity and beam parameters along the RFQ length was the SPP RFQ which is DEMIRCI's one of the very first applications. The plot containing the evolution of the cell length, beam energy, modulation, synchronous phase and radial aperture can be found in Fig.~\ref{fig:SANAEM-POP-RFQ}.

\subsection{Interoperability and Compatibility}
\indent
The interoperability between different programs is essential for both cross-check and continuity reasons. Especially when a new software library is introduced, the input-output file level compatibility ensures quick verification of the procedures and results. Additionally file level compatibility ensures the continuous processing of the ion beam from one end of the beamline to the other using different software, each suitable to the properties of a particular section. DEMIRCI was written with these goals in mind:~It can process input parameters produced by other programs and it produces output files compatible with most of the existing software packages in the field. 
For example, to compare with TOUTATIS, DEMIRCI produces a file which describes the ion source, ion properties, RFQ cells etc. 
The format of this file is described in the TOUTATIS documentation \cite{toutatis} and the values it contains are either
specified by the DEMIRCI user (e.g. ion mass and charge) or calculated by DEMIRCI library (e.g. RFQ cell definitions) 
or taken as TOUTATIS defaults (e.g. input and output matching section gap lengths).
The format of the section that defines the RFQ is the same as PARMETQM input file.
Although the TOUTATIS documentation specifies only 9 out of 16 parameters on this section are actually used,
all 16 columns are calculated by DEMIRCI and fed to TOUTATIS which then uses the input file to simulate the designed RFQ.
A different example would be CST Microwave Studio, for which there is no file exchange but the same cavity has to be defined from scratch using the programs own tools.
The compatibility between DEMIRCI and similar programs in the field, for basic parameters such as beam energy, RFQ length, quality factor and so on, is given in Table~\ref{tab:compatibility-from-Demirci's}. The input parameters used by different programs are all the same, originating from the previously discussed design of the SPP RFQ. As one may notice, the difference is usually of the order of few percent which can be considered as acceptable for a design program based solely on two term potential. The results from different software libraries for calculating the parameters along the RFQ are discussed below.

\begin{table}[!htb]
\caption{Results from DEMIRCI's Calculations as Compared to Other Programs Results (Keys are T:TOUTATIS, L:LIDOS, C:CST, S:SUPERFISH)\label{tab:compatibility-from-Demirci's} }

~

\centering{}%
\begin{tabular}{r@{\extracolsep{0pt}.}l|c|r|r}
\multicolumn{2}{c|}{\textbf{Parameter}} & \textbf{DEMIRCI} & \textbf{Other} & \textbf{\%$\Delta$}\tabularnewline
\hline 
\hline 
\multicolumn{2}{c|}{RFQ Length (m)} & 1.555 & 1.585 {[}L{]} & 1.89\tabularnewline
\cline{4-5} 
\multicolumn{2}{c|}{} &  & 1.549 {[}T{]} & 0.39\tabularnewline
\hline 
\multicolumn{2}{c|}{Exit Energy (MeV)} & 1.54 & 1.52 {[}L{]} & 1.32\tabularnewline
\cline{4-5} 
\multicolumn{2}{c|}{} &  & 1.49 {[}T{]} & 3.36\tabularnewline
\hline 
\multicolumn{2}{c|}{Travel Time (ns)} & 249.9 & 265.8 {[}L{]} & 5.98\tabularnewline
\cline{4-5} 
\multicolumn{2}{c|}{} &  & 243.8 {[}T{]} & 2.50\tabularnewline
\hline 
\multicolumn{2}{c|}{Quality Factor} & 10477.7 & 10341.6 {[}S{]} & 1.32\tabularnewline
\cline{4-5} 
\multicolumn{2}{c|}{} &  & 10216.4 {[}C{]} & 2.56\tabularnewline
\hline 
\multicolumn{2}{c|}{RF Power (W/cm)} & 122.23 & 123.56 {[}S{]} & 1.08 \tabularnewline
\cline{4-5} 
\multicolumn{2}{c|}{} &  & 125.08 {[}C{]} & 2.28\tabularnewline
\end{tabular}
\end{table}

\subsubsection{DEMIRCI-LIDOS}
\begin{figure}[!htb]
\begin{centering}
\includegraphics[bb=0bp 0bp 567bp 384bp,width=1\columnwidth]{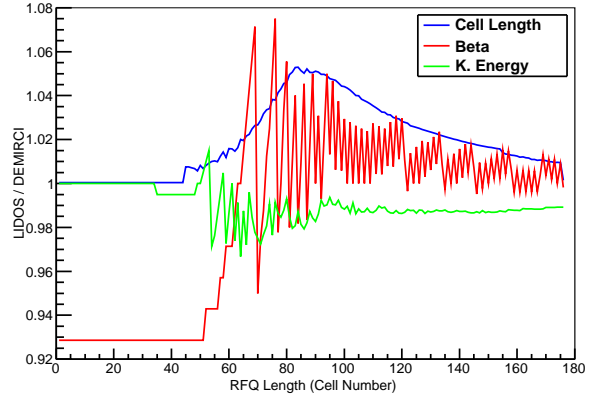}
\par\end{centering}
\caption{Comparisons of the DEMIRCI and LIDOS results for main parameters.\label{fig:Demirci-and-Lidos}
\label{fig:Demirci-and-Lidos-kine}}
\end{figure}

\begin{figure}[!htb]
\begin{centering}
\includegraphics[bb=0bp 0bp 567bp 384bp,width=1\columnwidth]{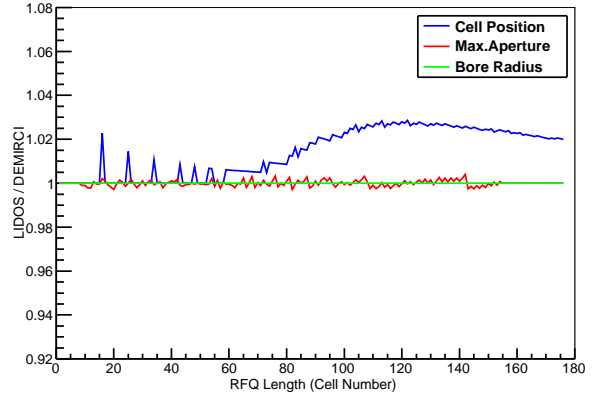}
\par\end{centering}
\caption{Design geometry comparisons of DEMIRCI and LIDOS. \label{fig:Demirci-Lidos-Geo}}
\end{figure}

\indent
The LIDOS.RFQ.Designer software package has three modules: one for design, one for analysis and one for simulations. 
The results used to compare this software to DEMIRCI are obtained from the first module, which also uses an 8-term potential to calculate the RFQ parameters. 
Fig.~\ref{fig:Demirci-and-Lidos} and Fig.~\ref{fig:Demirci-Lidos-Geo} contain the ratios of LIDOS results to DEMIRCI calculations for various variables such as kinetic energy and cell position. 
These results are obtained by uniquely defining the same RFQ in each program with the same four parameters (Synchronous Phase, Intervane Voltage, Minimum Bore Radius and Modulation).
The maximum of 3-5\% discrepancy between LIDOS and DEMIRCI results is assumed to arise from different calculation procedures (e.g. 8-term vs 2-term potential).

\subsubsection{DEMIRCI-TOUTATIS}

\indent

\begin{figure}[!htb]
\begin{centering}
\includegraphics[width=1\columnwidth]{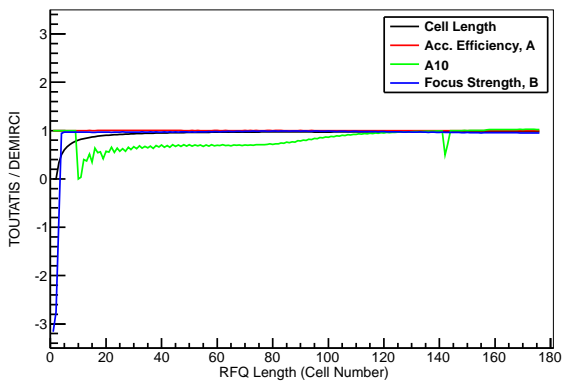}
\par\end{centering}
\caption{ Comparisons of the results from DEMIRCI and TOUTATIS. \label{fig:Demirci-and-Toutatis}}
\end{figure}

\textit{\emph{TOUTATIS}}\emph{ }\textit{\emph{\cite{toutatis} is a beam dynamics simulator for high intensity RFQs and makes this using Poisson solver for fields which provides to get values such as space charge effects or cavity fields between the electrodes. }}It interacts with DEMIRCI for RFQ beam dynamics simulation purposes. It describes an RFQ beamline with an input file containing different sections. The section that deals with the RFQ cells has a channel block definition compatible with PARMTEQM. DEMIRCI's default output file is compatible with this channel block definition containing RFQ cell parameters (inter-vane voltage, aperture, synchronous phase etc.) given in 17 columns. In fact, DEMIRCI's current version relies on TOUTATIS for all beam dynamics simulations. Fig.~\ref{fig:Demirci-and-Toutatis} shows a comparison of the main parameters obtained from DEMIRCI and TOUTATIS. It should be noted that the negative ratio observed in the first few cells of the Focusing Strength (B) curve
originates from TOUTATIS (v1.3), which reports negative values for these cells. This issue is currently under investigation. 

\begin{figure}[!htb]
\begin{centering}
\includegraphics[scale=0.5]{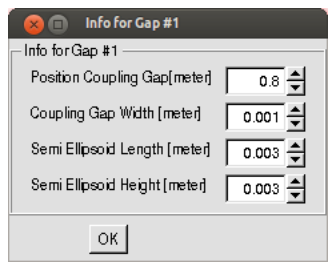}
\par\end{centering}
\caption{The RFQ segment gap definitions window. \label{fig:gap}}
\end{figure}

An important parameter is the RFQ length which is typically in the order of few meters. However, with today's CNCs, the optimal machining length is about 80-90~cm. Therefore RFQs are designed to be built and assembled in longitudinal sections. As the gap that occurs between different sections affects directly the behaviour of the ion beam inside the RFQ, it is important to incorporate it into the beam dynamics simulations. Although DEMIRCI doesn't directly use the section gaps for its calculations, it is possible to define any number of gaps and associated properties (such as location, width, etc.) as shown in Fig.~\ref{fig:gap} and to estimate the beam dynamics effects by running the design in TOUTATIS.

\subsubsection{LANL software suite}
\indent
\textit{\emph{LANL software suite for RFQ design is perhaps one of the most commonly used computer codes which can take an input file from DEMIRCI. SUPERFISH }}\cite{superfish}\textit{\emph{ is the most common computer program to calculate fields in 2D cartesian coordinates for any cylindrically symmetric shape. As SUPERFISH is an essential tool with years of experience, it is inevitable that DEMIRCI, professing being one of the standard computer codes, can produce SUPERFISH compatible output given an RFQ design. On the other hand, PARMTEQM is a common program that designs and makes beam dynamics calculating up to 8 term potential.}}

\section{Conclusions and Prospects}
\indent
DEMIRCI is a fast, Unix based modern tool using graphical techniques for RFQ design. It uses analytical formulae based on two term potential to compute the light ion beam behaviour in an RFQ. It also permits the user to achieve optimizations with specific goals such as a final accelerator length or a final ion beam energy. It interacts with similar software in the field, for result cross check and for further study of the RFQ and beam properties. Lastly, it can produce the horizontal and vertical vane shapes which can be fed into 3D solvers for more accurate electromagnetic and thermal studies based on finite element analysis techniques.

A number of additions and enhancements are being planned for this new tool. The first goal is to use the more complex 8 term potential to allow a more realistic calculation of the EM fields inside the RFQ. This enhancement is expected to further reduce the small deviations in the results obtained with this tool and similar ones. Apart from the focusing strength and defocusing factor, the inclusion of the parameters related to the stability of the transverse motion (e.g. phase advance and dispersion), is ongoing. The authors aim to provide various checks on the beam stability in the next version of the program. Furthermore, addition of beam dynamics calculations would make DEMIRCI a more complete solution for the RFQ design.

\section*{Acknowledgments}
\indent
The authors are grateful to S. Sultansoy, O. Cakir for useful comments and Jim Stovall for fruitful discussions. The authors are also grateful to TAEK for their endorsements and to LANL for their encouragements. This study is supported in part by the TAEK project under grant No.~A1.H4.P1.03. 

\section*{Appendix A}\label{apa}
In the classical $Q$ factor calculation, the electric field is not taken into consideration as its tangential component vanishes at the cavity wall~\cite{power_E}.
Its longitudinal component appears only in the presence of modulated vanes and moreover inside small volume defined by
the bore radius and the cell length.
As an improvement to eqn.(\ref{oldQ}), we propose taking into account the modulation effect as follows.
Let the power term originating from the $z$ component of the electric field, which in turn is due to the vane modulation,
be denoted as  $P_{\ell}^{v}$. It can be calculated by adapting the dielectric loss formula, although the power will not be 
lost since the medium is not a dielectric but vacuum ~\cite{cst}:
\begin{equation}
P^{v} \equiv \pi f \epsilon_0 \int \left|  E_z \right|^{2}dv  \quad,
\end{equation}
\noindent where $\epsilon_{0}$ is the vacuum permittivity, ${E_z}$ is the electric field in $z$ direction and 
 ($dv$) is the integration volume. 
 The axial  electric field  integrated at the electrode tips and averaged over a cell length ($L/2$) becomes
independent of the surface element:
\begin{equation}
E_0 = ￼A V / L \quad ,
\end{equation}
where all variables have an implicit $z$ dependence making these cell number dependent, suitable for a cell by cell computation.
The integration volume ($dv$) is assumed to be a right square prism of sides  $r_0\times r_0 \times L/2$ for each cell.
Therefore replacing the integral by a summation formula, the new term becomes:
\begin{equation}
P^{v}={1\over 2} w \epsilon_{0} \sum_n (E_{0}^{2} {L \over 2} r_0 ^2) \quad,
\end{equation}
where the summation is over all RFQ cells. The electric energy per unit length due to modulated vanes can be also calculated similarly:
\begin{equation}
U_{\ell}^{E}\equiv \frac{\epsilon_{0}}{2}\int\left|\bf{E}\right|^{2}dS = {1\over 2}  \epsilon_{0} \sum_n (E_{0}^{2} r_0 ^2) \quad,
\end{equation}
Therefore we propose a refined quality factor defined as:
\begin{equation}
Q=\omega\frac{U^{E}+U^{B}}{P^{s}+P^{v}}\quad,
\end{equation}
\noindent where terms without the subscript $\ell$ mean integration over the RFQ length. 
Typical contribution from $U^E$ is about 5 percent of the $U^B$, and from $P^{v}$ is about 8 percent of $P^{s}$.
The new formula reduces the over optimistic quality factor in eqn.(\ref{oldQ}) by about 2-3 percent. The choice between these
two reported quality factor values is left to the user.

\end{document}